\documentclass[twocolumn]{aastex631}
\usepackage{graphicx,xspace,apjfonts}

\newcommand\aql{31\,Aql\xspace}
\newcommand\xray{\mbox{X-ray}\xspace}
\newcommand\Rocrit{\mathrm{Ro_{crit}}}
\newcommand\Roflip{\mathrm{Ro_{flip}}}
\newcommand\Rosun{\mathrm{Ro_\odot}}
\shorttitle{Subgiant 31\,Aquilae}
\shortauthors{Metcalfe et al.}

\journalinfo{The Astronomical Journal (accepted version)}

\begin{document}

\title{\large Anti-Solar Differential Rotation May Have Revived Magnetic Braking in the Subgiant 31\,Aquilae}

\author[0000-0003-4034-0416]{Travis S.~Metcalfe}
\affiliation{Center for Solar-Stellar Connections, WDRC, 9020 Brumm Trail, Golden, CO 80403, USA}

\author[0000-0002-4284-8638]{Jennifer L.~van~Saders}
\affiliation{Institute for Astronomy, University of Hawai`i, 2680 Woodlawn Drive, Honolulu, HI 96822, USA}

\author[0000-0002-1242-5124]{Thomas R.~Ayres}
\affiliation{Center for Astrophysics and Space Astronomy, 389 UCB, University of Colorado, Boulder, CO 80309, USA}

\author[0000-0002-1988-143X]{Derek Buzasi}
\affiliation{Department of Astronomy \& Astrophysics, University of Chicago, 5640 S. Ellis Avenue, Chicago, IL 60637, USA}

\author[0000-0002-0210-2276]{Jeremy J.~Drake}
\affiliation{Lockheed Martin Solar and Astrophysics Laboratory, 3251 Hanover St, Palo Alto, CA 94304, USA}

\author[0000-0002-4996-0753]{Ricky Egeland}
\affiliation{Center for Solar-Stellar Connections, WDRC, 9020 Brumm Trail, Golden, CO 80403, USA}

\author[0000-0002-8854-3776]{Rafael A.~Garc\'\i a}
\affiliation{Universit\'e Paris-Saclay, Universit\'e Paris Cit\'e, CEA, CNRS, AIM, 91191, Gif-sur-Yvette, France}

\author[0000-0003-3061-4591]{Oleg Kochukhov}
\affiliation{Department of Physics and Astronomy, Uppsala University, Box 516, SE-75120 Uppsala, Sweden}

\author[0000-0001-7032-8480]{Steven H.~Saar}
\affiliation{Harvard-Smithsonian Center for Astrophysics, Cambridge, MA 02138, USA}

\author[0000-0002-3481-9052]{Keivan G.~Stassun}
\affiliation{Vanderbilt University, Department of Physics \& Astronomy, 6301 Stevenson Center Lane, Nashville, TN 37235, USA}

\author[0000-0002-6163-3472]{Sarbani Basu}
\affiliation{Department of Astronomy, Yale University, PO Box 208101, New Haven, CT 06520-8101, USA}

\author[0000-0001-7664-648X]{J.~M.~Joel Ong}
\affiliation{Sydney Institute for Astronomy (SIfA), School of Physics, University of Sydney, Camperdown, NSW 2006, Australia}

\author[0000-0002-5496-365X]{Amalie Stokholm}
\affiliation{School of Physics \& Astronomy, University of Birmingham, Edgbaston, Birmingham B15 2TT, UK}

\author[0000-0001-5222-4661]{Timothy R.~Bedding}
\affiliation{Sydney Institute for Astronomy (SIfA), School of Physics, University of Sydney, Camperdown, NSW 2006, Australia}

\author[0000-0003-0377-0740]{Sylvain N.~Breton}
\affiliation{INAF – Osservatorio Astrofisico di Catania, Via S. Sofia, 78, 95123 Catania, Italy}

\author[0000-0002-0551-046X]{Ilya V.~Ilyin}
\affiliation{Leibniz-Institut f\"ur Astrophysik Potsdam (AIP), An der Sternwarte 16, D-14482 Potsdam, Germany}

\author[0000-0001-7624-9222]{Pascal Petit}
\affiliation{Universit\'e de Toulouse, CNRS, CNES, 14 avenue Edouard Belin, 31400, Toulouse, France}

\author[0000-0002-7549-7766]{Marc H.~Pinsonneault}
\affiliation{Department of Astronomy, The Ohio State University, 140 West 18th Avenue, Columbus, OH 43210, USA}

\author[0000-0002-6192-6494]{Klaus G.~Strassmeier}
\affiliation{Leibniz-Institut f\"ur Astrophysik Potsdam (AIP), An der Sternwarte 16, D-14482 Potsdam, Germany}

\begin{abstract}

Recent observations have shown that sufficiently slow rotation disrupts the organization 
of large-scale magnetic field in older main-sequence stars, leading to weakened magnetic 
braking (WMB) and a collapse in the efficiency of the global stellar dynamo. Recent 
simulations predict a shift from solar-like to anti-solar differential rotation (DR) at 
slower rotation rates, which typically do not occur on the main-sequence due to WMB. 
However, physical expansion on the subgiant branch can eventually slow the stellar 
rotation beyond this threshold, yielding a non-cycling large-scale field that revives 
magnetic braking. We combine asteroseismology from the Transiting Exoplanet Survey 
Satellite (TESS) with spectropolarimetry from the Large Binocular Telescope (LBT) to test 
these predictions in the old metal-rich subgiant \aql. The LBT observations reveal a 
strong large-scale magnetic field in this star, and archival measurements of its 
chromospheric emission over 50 years confirm that it is non-cycling, as predicted. The 
star exhibits a variety of rotation periods during different observing seasons, 
consistent with DR but with no means of distinguishing between solar-like and anti-solar 
patterns. We incorporate the TESS observations to estimate the current wind braking 
torque of \aql, demonstrating that it supports revived magnetic braking in this old 
subgiant. We also use rotational evolution modeling to place a preliminary constraint on 
the stellar Rossby number for the transition to anti-solar DR. Future refinements in both 
asteroseismic observations and rotational modeling may yield improvements to this initial 
analysis.

\end{abstract}


\section{Introduction}\label{sec1}

The influence of stellar rotation on global convective patterns changes dramatically 
throughout the lives of solar-type stars. In the earliest phases of stellar evolution, 
rapid rotation is thought to imprint a cylindrical pattern of differential rotation (DR) 
on the outer convection zone, with a fast equator and slower poles in a configuration 
known as the Taylor-Proudman state \citep{Proudman1916, Taylor1917}. As the magnetized 
stellar wind gradually sheds angular momentum, rotation slows and the DR pattern is 
thought to become more conical as observed in the Sun \citep{Thompson1996}, with contours 
of constant rotation extending almost radially through the convection zone rather than 
parallel to the rotation axis. When the rotation rate becomes comparable to the 
convective overturn timescale ($\tau_c$), a broad array of convection simulations suggest 
that the DR pattern will flip from solar-like to anti-solar, with a slow equator and 
faster poles \citep{Gastine2014}. However, observational confirmation of anti-solar DR 
has been mostly confined to red giant stars \citep{Weber2005}.

Our understanding of stellar rotational evolution has been updated substantially over the 
past decade, modifying our expectations for the detection of anti-solar DR. After stars 
contract onto the main-sequence and magnetic braking begins to dominate their rotational 
evolution, the empirical relations of \cite{Skumanich1972} approximate the decline of 
stellar activity and the gradual slowing of rotation with the square-root of stellar age. 
The discovery of weakened magnetic braking \citep[WMB;][]{vanSaders2016} in old Kepler 
field stars was the first indication that this orderly progression does not continue 
indefinitely, but is disrupted at activity levels close to solar \citep{Metcalfe2016}. 
Observational constraints on the large-scale magnetic field strength and the mass-loss 
rates of older stars revealed unexpected changes in both properties that combine to 
produce WMB and keep stellar rotation nearly constant during the second half of 
main-sequence lifetimes \citep{Metcalfe2025b}. Consequently, stars may not reach the slow 
rotation rates that are required for the shift to anti-solar DR until the physical 
expansion associated with core H exhaustion finally pushes them over this threshold.

Recent simulations predict some observational signatures that accompany the shift from 
solar-like to anti-solar DR, including the emergence of a non-cycling large-scale 
magnetic field. A set of dynamo simulations described by \cite[][their Fig.13]{Brun2022} 
appear to show a gradual transition from short activity cycles for rapidly rotating 
models to longer sun-like cycles for intermediate rotation near the solar rate, and no 
cycles for models rotating more slowly than the Sun. Similar results were reported by 
\cite{Strugarek2017} for a completely independent set of dynamo simulations, and both 
sets of models lend qualitative support to the evolutionary scenario proposed by 
\cite{Metcalfe2017}. The simulations also appear to show a gradual decline in the 
large-scale magnetic field strength as activity cycles grow longer, reaching a minimum 
near the solar rotation rate and increasing once again for more slowly rotating models 
with no cycles and anti-solar DR (\citeauthor{Noraz2024}\,\citeyear{Noraz2024}; see also 
\citeauthor{Karak2015}\,\citeyear{Karak2015}, 
\citeauthor{Brandenburg2018}\,\citeyear{Brandenburg2018}). These results suggest that 
stars with anti-solar DR might be expected to show ``flat activity'' from a stationary 
large-scale magnetic field. Crucially, the rotation rates that are required to produce 
anti-solar DR in the simulations appear to be slower than FGK stars can typically reach 
on the main-sequence under the influence of WMB.

The metal-rich G7 subgiant \aql (b\,Aql, HD\,182572, TIC\,359981217) may have already 
evolved through the transition to anti-solar DR. In more than 50 years of chromospheric 
activity monitoring from the Mount Wilson survey \citep{Baliunas1995} and the Keck 
observatory \citep{Baum2022}, \aql shows minimal long-term variability around a low 
average activity level ($\log R'_{\rm HK}=-5.1$), while the short-term variability is 
sufficient to measure a rotation period of $P_{\rm rot}\sim41$~days \citep{Baliunas1996}. 
The latter suggests a Rossby number ($\mathrm{Ro} \equiv P_{\rm rot}/\tau_c$) that may be 
well above the solar value, potentially in the realm of anti-solar DR. In 
Section~\ref{sec2} we analyze new and archival observations to characterize \aql, 
including time series photometry from the Transiting Exoplanet Survey Satellite 
\citep[TESS;][]{Ricker2014}, spectropolarimetry from the Large Binocular Telescope (LBT), 
\xray measurements from Chandra, and ultraviolet observations from Hubble. In 
Section~\ref{sec3} we use these observations to determine precise stellar properties from 
asteroseismology, and to constrain the evolutionary pathway that led to the current 
magnetic and rotational configuration. Finally, in Section~\ref{sec4} we discuss our 
results and compare them to the predictions of dynamo simulations, to assess the 
possibility that \aql might exhibit anti-solar DR.

\section{Observations}\label{sec2}

Below we analyze new and archival observations of \aql to characterize the stellar 
properties that are relevant to its magnetic and rotational evolution. In 
Section~\ref{sec2.1} we identify solar-like oscillation frequencies from recent TESS 
photometry. In Section~\ref{sec2.2} we analyze new LBT spectropolarimetry to estimate the 
large-scale magnetic field strength, and we verify its persistence in archival 
observations. In Section~\ref{sec2.3} we use recent \xray measurements from Chandra to 
estimate the mass-loss rate, and in Section~\ref{sec2.4} we present new ultraviolet 
observations from Hubble. In Section~\ref{sec2.5} we use archival Mount Wilson data to 
confirm the absence of an obvious activity cycle and to extract seasonal rotation 
measurements. Finally, in Section~\ref{sec2.6} we determine the bolometric luminosity 
from the spectral energy distribution (SED).

\subsection{TESS Photometry}\label{sec2.1}

\begin{figure}[t]
 \centering\includegraphics[width=\columnwidth]{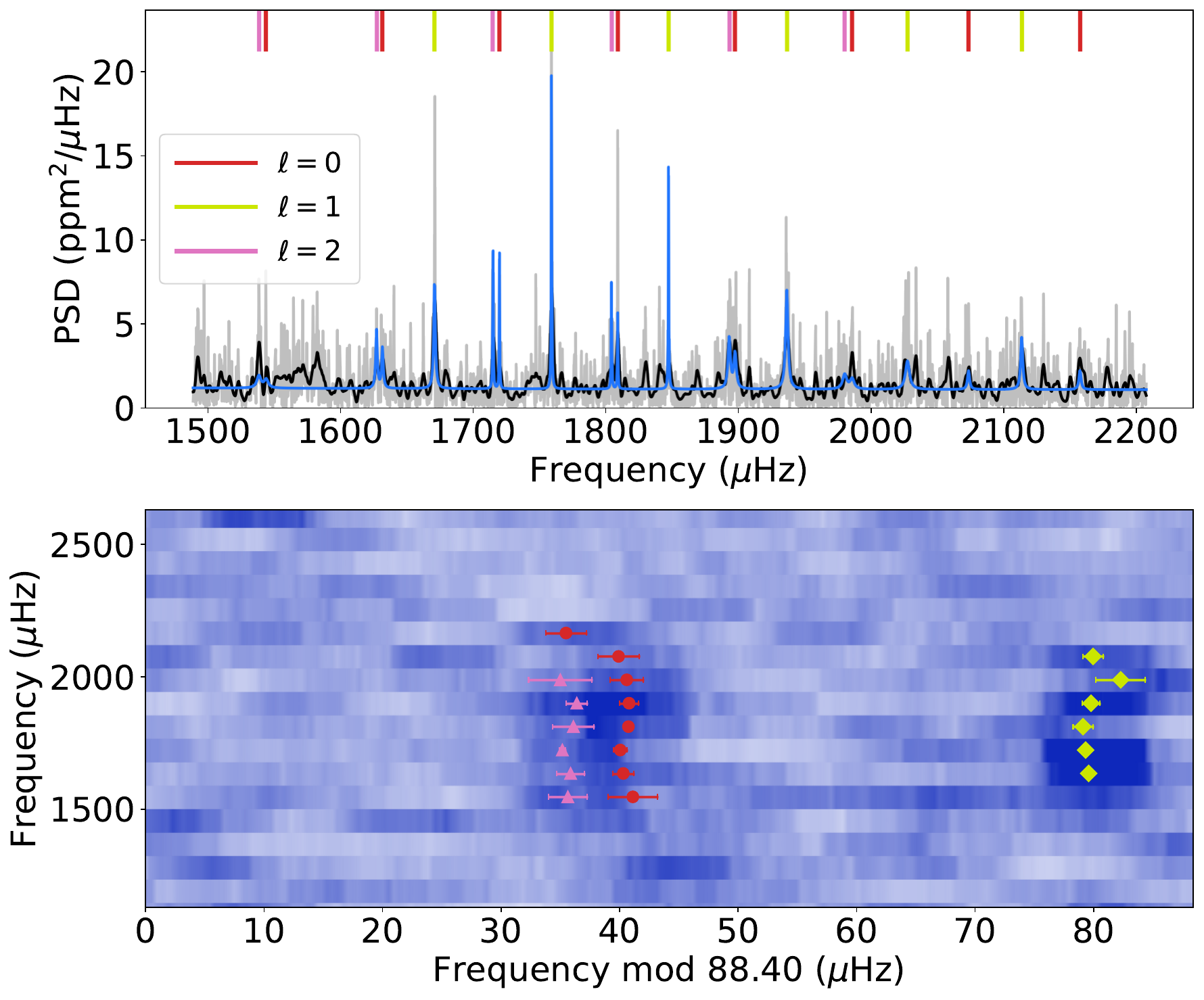}
 \caption{{\it Top:} Power spectral density (PSD) in the frequency range of the fitted modes. 
 In gray is the raw spectrum and in black a smoothed PSD. The blue line represents the 
fitted spectrum. The vertical red, yellow, and magenta bars indicate the central 
frequencies of the $\ell=0$, 1, and 2 modes, respectively. {\it Bottom:} \'Echelle 
diagram with \mbox{$\Delta \nu=88.40$ $\mu$Hz}. The fitted modes and their associated 
uncertainties are shown as circles, diamonds, and triangles with the same color code as 
in the top panel.\label{fig1}}
 \end{figure}

TESS observed \aql at 120~s cadence during Sector 54 \citep[see][]{Lund2025} and at 20~s 
cadence during Sector 81 (2024 July 15 -- 2024 August 10). In this analysis we use only 
the Sector 81 data because the 20~s cadence has lower noise than 120~s cadence data for 
bright stars \citep{Huber2022}. We extracted the data from the raw image frames following 
the process described in \cite{Nielsen2020} and \cite{Metcalfe2023b}, which uses aperture 
masks optimized to produce the highest signal-to-noise (S/N) ratio possible with simple 
aperture photometry on an isolated star; in this case our mask contained 85 pixels from 
the 297-pixel postage stamp. Conservatively, we discarded observations with flag values 
greater than unity, leaving 106,244 data points with a duty cycle above 92\%. Finally, we 
detrended the light curve against centroid pixel coordinates, breaking the time series 
into two parts separated by the data downlink gap starting near day 3519. The resulting 
light curve has a noise level that is approximately 11\% lower compared to the Science 
Processing Operations Center \citep[SPOC;][]{Jenkins2016} PDCSAP product. Prior to 
frequency analysis, the light curve was gap filled using a multi-scale discrete cosine 
transform following inpainting principles \citep{Garcia2014a, Pires2015}, and high-pass 
filtered with a cutoff frequency of $100~\mu$Hz to minimize residual contributions from 
spacecraft jitter. The power spectrum of this light curve is illustrated in the top panel 
of Figure~\ref{fig1}, with an \'echelle diagram in the bottom panel to facilitate the 
identification of the spherical harmonic degree ($\ell$) for each frequency.

The solar-like oscillation frequencies were identified using the 
\texttt{apollinaire}\footnote{Documentation is available at 
\url{https://apollinaire.readthedocs.io/}.} code \citep{Breton2022}, which is built upon 
the \texttt{emcee} ensemble Markov Chain Monte Carlo sampler \citep{ForemanMackey2013}. 
In this framework, each mode is described by a Lorentzian function, where both the 
central frequency and the linewidth are treated as independent parameters. Rather than 
fitting an individual height for each mode, a single height parameter is assigned ``by 
order $n$'', i.e., to all modes with $\ell = 0, 1$ at a given radial order $n$, and to 
$\ell = 2$ modes at order $n-1$. The ratios of mode heights between different angular 
degrees are incorporated as global parameters. The free parameters for all modes were fit 
simultaneously in a global procedure, following the methodology pioneered by 
\citet{RocaCortes1999} and improved by \citet{Fletcher2008}, which helps to account for 
correlations between modes and provides a more robust estimation of the oscillation 
parameters \citep[see also][]{Davies2016, Lund2017}. The identified oscillation 
frequencies and their uncertainties are listed in Table~\ref{tab1} and shown in 
Figure~\ref{fig1}. Posterior probability distributions were generated for each parameter, 
and the median of each distribution was adopted as the central value. The adopted 
uncertainties are the average of the differences between the median and the 16th and 84th 
percentiles of the posterior distributions.

 \begin{deluxetable}{llcc}
 \setlength{\tabcolsep}{20pt}
 \tablecaption{Identified Oscillation Frequencies for \aql \label{tab1}}
 \tablehead{ \colhead{$n$} & \colhead{$\ell$} & \colhead{$\nu$ ($\mu$Hz)} & \colhead{$\sigma_\nu$ ($\mu$Hz)} }
 \startdata
 17 & 0     & 1543.92 & 2.09 \\
 18 & 0     & 1631.51 & 0.91 \\
 19 & 0     & 1719.66 & 0.57 \\
 20 & 0     & 1808.74 & 0.20 \\
 21 & 0     & 1897.20 & 0.81 \\
 22 & 0     & 1985.40 & 1.39 \\
 23 & 0     & 2073.11 & 1.75 \\
 24 & 0$^*$ & 2157.08 & 1.71 \\
 18 & 1     & 1670.78 & 0.38 \\
 19 & 1     & 1758.92 & 0.17 \\
 20 & 1     & 1847.10 & 0.83 \\
 21 & 1     & 1936.18 & 0.75 \\
 22 & 1     & 2027.08 & 2.08 \\
 23 & 1     & 2113.15 & 0.83 \\
 16 & 2     & 1538.43 & 1.65 \\
 17 & 2     & 1627.07 & 1.16 \\
 18 & 2     & 1714.77 & 0.28 \\
 19 & 2     & 1804.10 & 1.74 \\
 20 & 2     & 1892.79 & 0.87 \\
 21 & 2     & 1979.80 & 2.67 \\
 \enddata
 \tablecomments{$^*$ Possible misidentification of an $(n=23,\ \ell=2)$ mode.}
 \end{deluxetable}
 \vspace*{-12pt}

\subsection{Spectropolarimetry}\label{sec2.2}

We observed \aql with the Potsdam Echelle Polarimetric and Spectroscopic Instrument 
\citep[PEPSI;][]{Strassmeier2015} installed at the $2\times8.4$~m LBT. The observation 
comprised a single snapshot taken on 2025 July 6. We employed the same instrumental setup 
(spectral resolution $R=130,000$ with wavelength coverage spanning the 475--540 and 
623--743~nm regions) and data reduction procedures as described in \citet{Metcalfe2019}. 
Given the low expected amplitude of the polarization signal, we applied the least-squares 
deconvolution \citep[LSD;][]{Donati1997, Kochukhov2010} technique to extract high S/N 
mean intensity and circular polarization profiles.

\begin{figure}[t]
 \centering\includegraphics[width=\columnwidth]{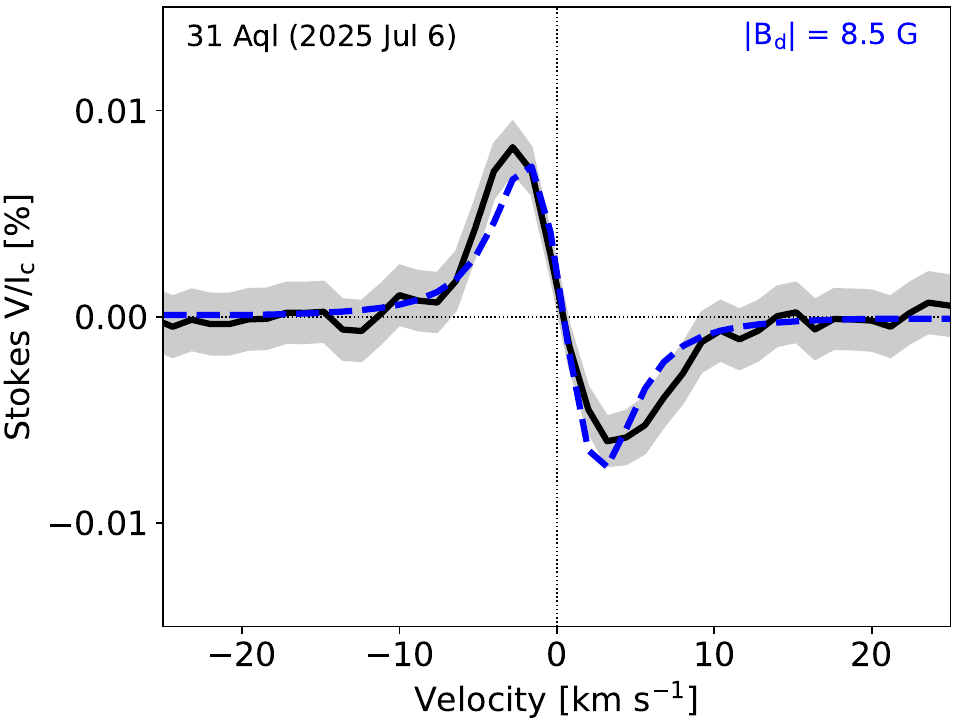} 
 \caption{Stokes~$V$ polarization profile for \aql from LBT observations on 2025~July~6. 
The observed LSD profile is shown as a black line, with uncertainties indicated by the 
gray shaded area. The dashed blue line is a model profile assuming an axisymmetric dipole 
morphology with a fixed inclination.\label{fig2}}
 \end{figure}

The line mask required for LSD was obtained from the Vienna Atomic Line Database 
\citep[VALD;][]{Ryabchikova2015}, with atmospheric parameters $T_{\rm eff}=5500$~K, $\log 
g=4.0$, and metallicity [M/H]$=+0.3$, comparable to \citet{Brewer2016}. For the analysis 
of the PEPSI observation, we used 2568 metal lines deeper than 10\% of the continuum. The 
LSD procedure yielded a polarization profile with a precision of 4.18~ppm and a clear 
detection of the stellar magnetic field, as illustrated in Figure~\ref{fig2}. The Stokes 
$V$ signature of \aql corresponds to a mean longitudinal magnetic field $\langle B_{\rm 
z} \rangle = 1.525\pm0.053$~G. For consistency with our previous analyses of Stokes~$V$ 
profiles, we estimated the strength of an axisymmetric dipole magnetic field using the 
modeling procedure described in \citet{Metcalfe2019}, fixing the inclination angle at 
$i=61\degr$ \citep{Bowler2023}. This analysis yielded a dipole field strength of $B_{\rm 
d}=8.5 \pm 1.1$~G. Due to geometric cancellation effects, adoption of an axisymmetric 
quadrupole or octupole yields a very strong magnetic field: $B_{\rm q}=133 \pm 13$~G or 
$B_{\rm o}=256 \pm 74$~G.

We complemented the PEPSI observation of \aql with an analysis of archival 
spectropolarimetric data retrieved from PolarBase \citep{Petit2014}. This database 
includes two ESPaDOnS observations of \aql obtained in 2007, as well as 15 NARVAL Stokes 
$V$ spectra acquired over three nights in 2019. Both spectropolarimeters cover the 
370--1000~nm wavelength range at a spectral resolution of $R=65,000$. All spectra were 
processed using the same LSD procedure described above, but employing a line mask 
consisting of 5,306 spectral lines deeper than 20\% of the continuum.

Among the 15 NARVAL observations, a definite Zeeman signature indicative of a global 
magnetic field was detected in every spectrum, whereas the two earlier ESPaDOnS 
observations yielded no detection owing to their lower S/N ratios. The LSD Stokes $V$ 
profiles of \aql obtained from NARVAL observations exhibit amplitudes of order 
$\sim10^{-4}$ of the continuum, similar to the LSD profile derived from the PEPSI 
spectrum, enabling measurements of the mean longitudinal magnetic field in the range of 
1--2~G with typical uncertainties of $0.3$~G. The circular polarization signal appears 
stable, with no significant variability detected between nights. To further quantify the 
strength of the large-scale magnetic field in these archival spectra of \aql, we selected 
the five NARVAL observations with the highest S/N ratios and applied the axisymmetric 
dipole fitting methodology described above. This analysis yielded an average dipole field 
strength of $B_{\rm d} = 7.9 \pm 0.7$~G, where the quoted uncertainty corresponds to the 
standard deviation of the fitting results for the individual spectra. Thus, the NARVAL 
measurements were consistent with those from PEPSI, suggesting long-term stability of the 
large-scale magnetic field.

\subsection{X-Ray Measurements}\label{sec2.3}

The subgiant \aql was captured in the ROSAT All-Sky survey in 1990 October as a faint 
source in 0.50~ks of accumulated exposure. The count rate was $0.029{\pm}0.009$ 
counts~s$^{-1}$ (0.1$-$2.4 keV) as reported in the rassfsc catalog (released circa 2000), 
and $0.037{\pm}0.011$ counts~s$^{-1}$ in the updated rass2rxs (circa 2016). More 
recently, in 2019 November, Chandra carried out a 9~ks pointing on \aql with the 
High-Resolution Camera (HRC-I: ObsID 22309; PI: T.~Metcalfe). Although HRC-I lacks any 
meaningful energy resolution, it was the only viable option for the soft coronal source 
\aql---the other Chandra camera (ACIS) is affected by an ongoing organic contamination 
issue that severely reduces the sensitivity to low-energy photons. Further, there were no 
observations of \aql listed in the archives of the other main contemporary high-energy 
facilities: XMM-Newton and the Swift \xray Telescope.

\begin{figure}[!t]
 \centering\includegraphics[width=\columnwidth]{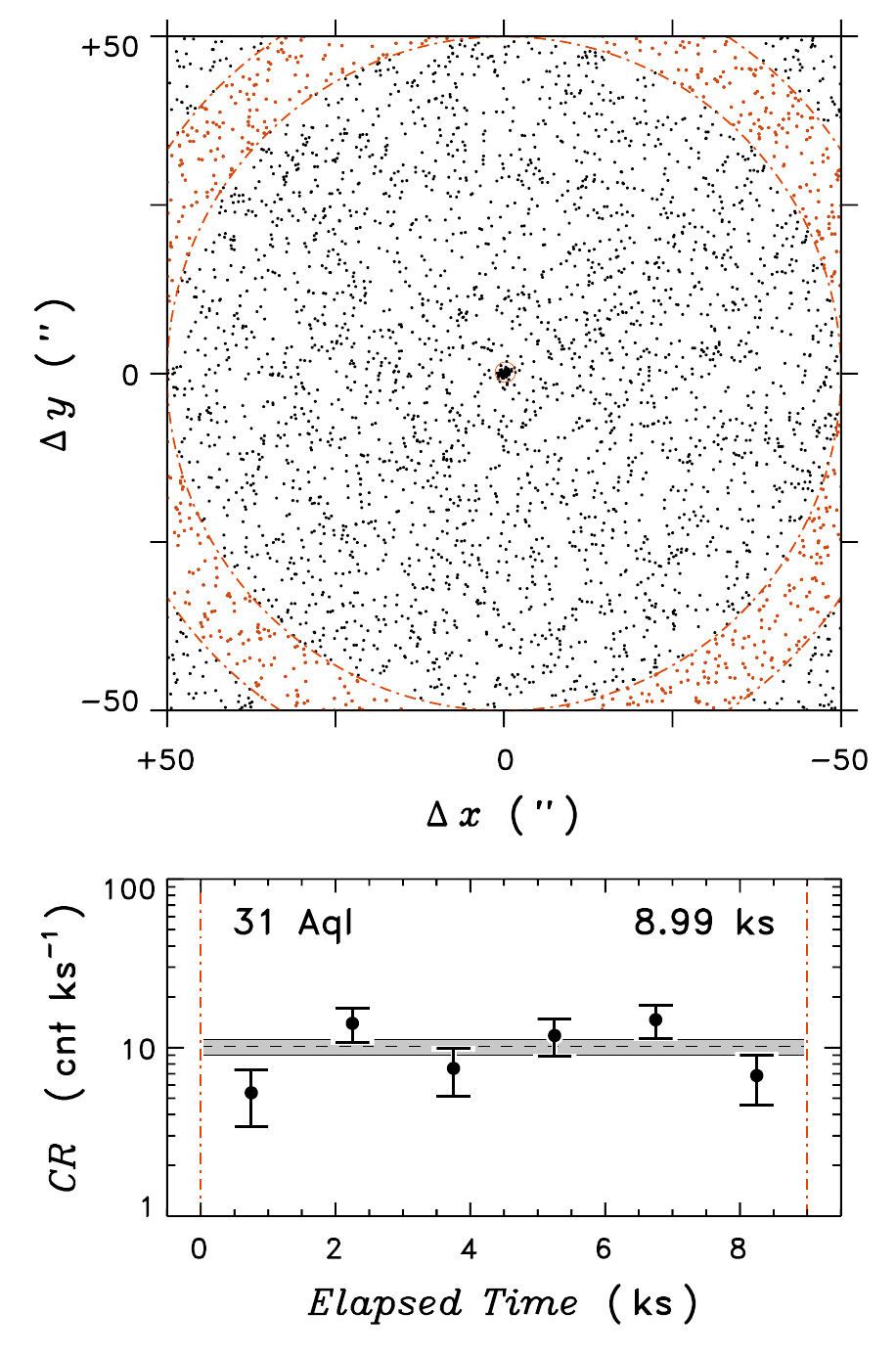}
 \caption{{\it Top:}\/ Photon map derived from the 2019 Chandra \mbox{HRC-I} pointing on 
\aql. The axes are sky coordinates relative to the source centroid. The small red circle 
is the 3{\arcsec} detection cell; the outer red circles, and red dots, delimit the 
background extraction zone. {\it Bottom:} Time-binned \xray count rates ($\Delta{t}= 
1500$s) from the HRC-I pointing, corrected for background and the encircled energy 
fraction, as a function of elapsed time (the dead-time corrected exposure was 8.99~ks). 
The gray band indicates the average count rate and $\pm$1\,$\sigma$ Poisson uncertainty 
for the full observation.\label{fig3}}
 \end{figure}

Figure~\ref{fig3} is a photon map, in relative sky coordinates, of the center of the 
HRC-I field, showing a prominent point source within 0\farcs4 of the predicted location 
of the bright star in that epoch (typical of the accuracy of the Chandra aspect 
solution). There are no other significant \xray point sources in the 
50{\arcsec}$\times$50{\arcsec} field. The small red circle indicates the adopted 
3{\arcsec}-diameter detection cell (95\% encircled energy). The Washington Double Star 
Catalog reports a faint visual component ``E'' to \aql (WDS J19250+1157A) in 2002 at 
position angle 288{\arcdeg}, separation of 4\farcs2, and a magnitude deficit of 
$\approx$7 in the K-band.  However, Gaia DR3 (epoch 2015.5) lists no entries within 
30{\arcsec} of the nearby subgiant, brighter than $G= 17$ or with a parallax greater than 
1~mas, so component E must have been a background object not sharing the high proper 
motion of \aql.

The \aql HRC-I level 2 event list from the Chandra Archive was processed as described by 
\cite{Ayres2025}. The gross number of counts in the 3{\arcsec}-diameter detection cell 
was 89${\pm}$9.4 in the 8.99~ks of dead-time corrected exposure. The background was 
sampled in an annulus 50{\arcsec}$-$60{\arcsec} from the source, resulting in an areal 
rate of $3.9{\times}10^{-5}$ counts~(\arcsec)$^{-2}$~s$^{-1}$, typical of HRC-I 
pointings. This implied a background contribution scaled to the detection cell of 2.5 
counts. The net HRC-I count rate, corrected for the encircled energy factor, was 
$0.0101{\pm}0.0011$ counts~s$^{-1}$. As shown in Figure~\ref{fig3}, there was no obvious 
variability of the source (e.g., flare activity) when the event list was blocked into 
multiple time segments (six was the maximum given the low count rate).

The ROSAT/PSPC and Chandra/HRC-I count rates were then calibrated into flux units by 
calculating optimum energy conversion factors (ECF) for the two instrument configurations 
based on a grid of coronal emission-measure models \citep[see][for details]{Ayres2025}. 
An apparent slow systematic decline of the HRC-I sensitivity since 2016 was taken into 
account. The ECF optimization made use of the adopted fundamental parameters of \aql, 
namely $T_{\rm eff}= $5587~K and $f_{\rm bol}= 2.24{\times}10^{-7}$ erg s$^{-1}$ 
cm$^{-2}$, and a hydrogen column density to the nearby star ($d= 14.92$~pc) of $N_{\rm 
H}= 1{\times}10^{18}$ cm$^{-2}$.

The ROSAT rass2rxs count rate yielded an \xray luminosity $L_{\rm X}= (4.5 \pm 1.3) 
\times 10^{27}$ erg s$^{-1}$, while the Chandra HRC-I result was somewhat lower, $L_{\rm 
X}= (2.4 \pm 0.3) \times 10^{27}$ erg s$^{-1}$.  The HRC-I luminosity is consistent with 
the less certain rass2rxs value, within the typical ranges of short-term and long-term 
variability seen in normal stars \citep{Ayres2025}. We adopt the higher-quality HRC-I 
\xray luminosity, recognizing that a factor of two uncertainty would not be unusual. 
However, as noted in Section~\ref{sec2.5}, the chromospheric activity level of the star 
was relatively constant between the epochs of the ROSAT and Chandra pointings.

We can estimate the mass-loss rate of \aql by combining the adopted \xray luminosity with 
the stellar radius determined from asteroseismology (see Section~\ref{sec3.1}). For stars 
with mass-loss rates inferred directly from observations of Ly$\alpha$, there is an 
empirical relation between the mass-loss rate and the \xray flux per unit surface area, 
$\dot M \propto F_{\rm X}^{1.29\pm0.16}$ \citep{Wood2018}. The resulting estimate is 
slightly above the solar value, $\dot{M} = 1.05^{+1.08}_{-0.58}\ \dot{M}_\odot$. The 
uncertainty accounts for the errors on $L_{\rm X}$, $R$ and the power law exponent 
combined in quadrature with a factor of 2 systematic uncertainty on the empirical 
relation \citep{Wood2005a}.

\subsection{Hubble Space Telescope Data}\label{sec2.4}

As a probe of the physical environment between the photosphere and the corona, we also 
observed \aql using the Cosmic Origins Spectrograph (COS) on the Hubble Space Telescope 
(HST). The description below closely follows a similar analysis of $\lambda$\,Ser by 
\cite{Metcalfe2023b}. We obtained a low resolution G140L FUV spectrum with an integration 
time of $\approx$1940~s on 2020 March 2 (program 15991), which was reduced with standard 
pipeline processing. We computed integrated fluxes for isolated emission lines in the 
rest frame of the star above nearby pseudo-continua, together with root-mean-square (RMS) 
errors. We estimated continua from linear fits to clusters of low points on either side 
of the line in question.  When the target line was blended, two methods were employed. 
Some weaker blends were removed by fitting a Voigt function to better mimic the convolved 
instrumental profile and line shape, leaving the residual target line for flux 
integration as before. In some cases, we fit the entire complex of lines with multiple 
Voigt functions. The results are listed in Table~\ref{tab2}. Tests on isolated lines 
demonstrated that straight integration and Voigt fitting yielded similar results, 
typically within $\pm$5\%. In several cases, multiple nearby lines of the same ion were 
combined. Following \citet{Ayres2020}, we also measured a 10~\AA\ segment of relatively 
line-free FUV pseudo-continuum centered at 1506~\AA.

 \begin{deluxetable}{lcc}
 \setlength{\tabcolsep}{13pt}
 \tablecaption{Measured FUV Line Fluxes for \aql \label{tab2}}
 \tablehead{ \colhead{Ion(s)} & \colhead{Wavelength} & \colhead{Flux at Earth} \\ 
 \colhead{}  &  \colhead{[\AA]}  & \colhead{[10$^{-15}$ ergs cm$^{-2}$ s$^{-1}$]}}
 \startdata
 C {\sc iii}$^1$            & 1175   & 10.1  $\pm$ 1.3  \\
 O {\sc i}$^1$              & 1304   & 55.4  $\pm$ 1.1  \\
 C {\sc ii}$^1$             & 1335   & 30.5  $\pm$ 0.9  \\
 Cl {\sc i}$^2$             & 1351.7 & 1.1   $\pm$ 0.3  \\
 O {\sc v}                  & 1371   & 0.21  $\pm$ 0.1  \\
 Si {\sc iv}                & 1393.8 & 8.0  $\pm$ 0.7   \\
 O {\sc iv}                 & 1399.8 & 0.10  $\pm$ 0.3  \\
 O {\sc iv}                 & 1401.2 & 0.48  $\pm$ 0.3  \\
 Si {\sc iv}$^2$            & 1402.8 & 5.5 $\pm$ 0.6    \\
 Si {\sc iv}+O {\sc iv}$^2$ & 1404.8 & 0.24  $\pm$ 0.6  \\
 O {\sc iv}$^2$             & 1407.4 & 0.21  $\pm$ 0.16 \\
 Continuum$^3$              & 1506   & 3.7  $\pm$ 0.8   \\
 Si {\sc ii}                & 1526.5 & 1.6   $\pm$ 0.9  \\ 
 Si {\sc ii}                & 1533.7 & 2.4   $\pm$ 0.8  \\ 
 C {\sc iv}$^2$             & 1548   & 15.3  $\pm$ 1.3  \\
 C {\sc iv}$^2$             & 1550   &  7.7  $\pm$ 1.1  \\
 C {\sc i}$^{1,2}$          & 1561   & 8.6  $\pm$ 1.0   \\ 
 He {\sc ii}                & 1640.7 & 6.2  $\pm$ 1.5   \\
 C {\sc i}$^1$              & 1657   & 29.9  $\pm$ 2.5  \\
 O {\sc iii}                & 1666.2 & 1.7   $\pm$ 1.2  \\
 Si {\sc ii}                & 1808.0 & 22.4  $\pm$ 3.1  \\ 
 Si {\sc ii}                & 1817.1 & 61.6 $\pm$ 5     \\ 
 Al {\sc iii}               & 1854.6 & 7.1  $\pm$ 3.6   \\
 Si {\sc iii}               & 1892.0 & 20.6  $\pm$ 6.6  \\
 C {\sc iii}                & 1908.7 & 11.8  $\pm$ 7.9  \\
 \enddata
 \tablecomments{All fluxes are from direct integration, except
  $^1$Multiple lines combined, $^2$Voigt function fitting and deblending,
  $^3\pm$5~\AA~integration of a largely line-free region.}
 \vspace*{-24pt}
 \end{deluxetable}
 \vspace*{-12pt}

We can compare the FUV fluxes of \aql to the Sun \citep{Ayres2020} and to $\lambda$\,Ser 
\citep{Metcalfe2023b}, a somewhat hotter solar analog (see Table~\ref{tab3}). With the 
exception of the hottest lines, \aql generally shows surfaces fluxes quite similar to the 
Sun.  Lower chromospheric surface flux in Cl~{\sc i} (with temperatures of peak 
emissivity $\log T_{\rm peak}$ = 3.8) is 1.3 times smaller than $\lambda$\,Ser; the upper 
chromospheric C~{\sc ii} flux ($\log T_{\rm peak} = 4.5$) is similarly $\approx 
1.5\times$ reduced. In hotter lines formed in the stellar transition region, C~{\sc iii} 
($\log T_{\rm peak} = 4.8$) is $\approx 1.9\times$ weaker relative to $\lambda$\,Ser, 
Si~{\sc iv} ($\log T_{\rm peak}$ = 4.9) is similarly reduced.  In the C~{\sc iv} doublet 
($\log T_{\rm peak}$ = 5.0), fluxes are somewhat stronger---a factor of $\approx$1.7 
lower than $\lambda$\,Ser, and 10\% weaker than the solar value (here some optical depth 
effects may play a role).

A number of density-sensitive line ratios can be found in the HST spectra.  We use a 
combination of these results to estimate the electron density in the stellar transition 
region. The ratio C~{\sc iii}(1908~\AA)/Si~{\sc iv}(1402~\AA), with $\log T_{\rm peak} 
\sim 4.8$, gives $\log n_e = 9.98_{-0.06}^{+0.05}$. The Si~{\sc iii}(1892~\AA)/C~{\sc 
iii}(1909~\AA) ratio, with $\log T_{\rm peak} \sim 4.7$, yields $\log n_e = 
9.87_{-0.18}^{+0.08}$ using \citet{Keenan1987}. The ratios O~{\sc iii}(1666~\AA)/Si~{\sc 
iv}(1402~\AA) and C~{\sc iii}(1908~\AA)/O~{\sc iii}(1666~\AA), also with $\log T_{\rm 
peak} \sim 4.8$, were less certain, affected by larger errors in the O~{\sc iii} line; 
these yield $\log n_e = 10.80_{-0.41}^{+0.40}$ and $9.76_{-1.44}^{+0.30}$, respectively 
\citep[using ][]{Keenan1988}.  Combining the diagnostics, we find an average $\langle 
\log n_e \rangle = 9.95 \pm 0.05$ at $\log T_{\rm peak} \sim 4.8$. For comparison, 
$\langle \log n_e \rangle = 10.0$ using the hotter O~{\sc iv} ratio in the Sun 
\citep[e.g.,][]{Rao2022}.  This suggests that \aql has transition region densities 
similar to, or perhaps slightly lower than, the Sun at fixed temperature, which is 
consistent with its lower surface gravity.  We caution, however, that assumptions 
intrinsic to the line ratio method make the results uncertain \citep[see discussion 
in][]{Judge2020}.

 \begin{deluxetable}{lcccc}
 \setlength{\tabcolsep}{7pt}
 \tablecaption{Comparison of FUV Surface Fluxes \label{tab3}}
 \tablehead{\colhead{Ion(s)} & \colhead{Wavelength} & \multicolumn{3}{c}{Surface Flux [10$^{3}$ ergs cm$^{-2}$ s$^{-1}$]} \\
 \colhead{}  &  \colhead{[\AA]}  & \aql & $\lambda$\,Ser & Sun$^2$ }
 \startdata    
 C {\sc iii}$^1$ & 1175   & 2290  &  4400  & 2250  \\
 O {\sc i}$^1$   & 1304   & 12600 &  8230  & 5490  \\
 C {\sc ii}$^1$  & 1335   & 6940  &  10700 & 7000  \\
 Cl {\sc i}      & 1351.7 &  245  &  330   &  252  \\
 Si {\sc iv}     & 1393.8 &  1830 &  4110  & 1690  \\
 Si {\sc iv}     & 1402.8 &  1260 &  1870  &  875  \\
 Continuum       & 1506   &  830  &  3210  & 1780  \\
 C {\sc iv}      & 1548   &  3490 &  6030  & 3800  \\
 C {\sc iv}      & 1550   &  1760 &  2730  & 1960  \\
 \enddata
 \tablecomments{$^1$Multiple lines combined, $^2$Results from \citet{Ayres2020}.} 
 \end{deluxetable}
 \vspace*{-12pt}

In summary, \aql has chromospheric and transition region fluxes broadly consistent with a 
star which has similar coronal activity, and has slightly lower density than the Sun.  
It is also a factor of $\sim$1.3--2 less active than the evolved solar analog 
$\lambda$\,Ser, with larger reductions towards hotter lines.

\subsection{Chromospheric Activity Data}\label{sec2.5}

 \begin{figure*}[t]
 \centering\includegraphics[width=\textwidth]{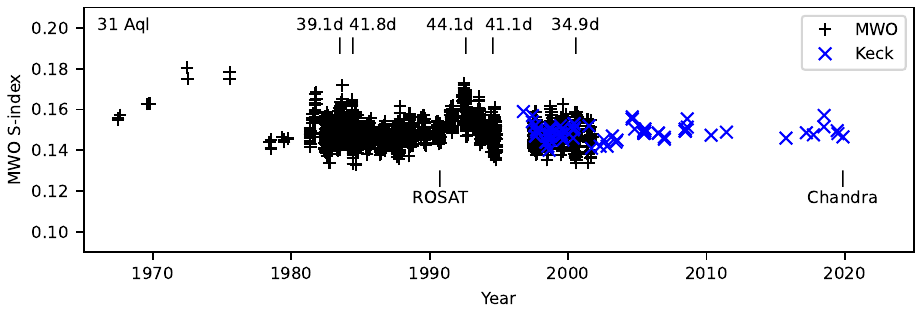}
 \caption{Chromospheric activity measurements of \aql spanning more than 50~years from 
Mount Wilson (black plus) and Keck (blue cross), showing minimal long-term variability 
but sufficient short-term variability to measure the rotation period. Data are from 
\cite{Baum2022}.\label{fig4}}
 \end{figure*}

The Mount Wilson Observatory (MWO) HK Project \citep{Wilson1978, Baliunas1995} defined 
the $S$-index of chromospheric activity based on the core of the Ca~{\sc ii} H \& K 
emission lines relative to nearby pseudo-continuum bands \citep{Vaughan1978}.  
\citet{Baum2022} combined these data with observations from Keck HIRES, and the composite 
time series is shown in Figure \ref{fig4}.  The combined time series shows more than 50 
years of relatively flat activity punctuated by a few periods of enhanced activity.  
Observations are sparse from 1966--1982, as well as during the Keck time series, but the 
period from 1982 through 2001 has a sufficient number of observations per season to 
attempt to measure a seasonal rotation period.  As described in \cite{Donahue1996}, the 
range of seasonal rotation periods is presumed to be due to surface features (spots or 
faculae) transiting the star at different rates due to DR.

 \begin{deluxetable}{lcccc}[b]
 \setlength{\tabcolsep}{7pt}
 \tablecaption{Seasonal Measurements from the MWO Data\label{tab4}}
 \tablehead{\colhead{Season} & \colhead{$N$} & \colhead{$\left< S \right>$} & \colhead{$P_{\rm rot}$ (days)} & \colhead{FAP (\%)}}
 \startdata
 1983 & 69 & $0.1523 \pm 0.0050$ & $39.1 \pm 0.6$ & 0.0056 \\
 1984 & 75 & $0.1499 \pm 0.0006$ & $41.8 \pm 0.4$ & 0.0006 \\
 1992 & 36 & $0.1599 \pm 0.0010$ & $44.1 \pm 0.5$ & 0.0556 \\
 1994 & 45 & $0.1486 \pm 0.0006$ & $41.1 \pm 3.0$ & 2.3 \\
 2000 & 25 & $0.1482 \pm 0.0012$ & $34.9 \pm 0.5$ & 2.6 \\
 \enddata
 \end{deluxetable}
 \vspace*{-12pt}

Using the methods described in \cite{Metcalfe2023b} we analyzed 14 out of 24 MWO seasons 
with $N \ge 20$ observations using the Lomb-Scargle periodogram in a Monte Carlo fashion 
to estimate the seasonal rotation period, false alarm probability (FAP), and uncertainty 
in period.  We detected significant seasonal rotation periods (FAP < 5\%; 95\% 
confidence) in 6 out of 14 seasons.  The results are listed in Table~\ref{tab4}, along 
with the seasonal mean $S$-index.  Five of the six significant periods were detected near 
times of high activity; there are local maxima in the seasonal mean time series in 1983, 
1989, 1992, and 1999.  This is to be expected as the $S$-index is enhanced by the 
presence of magnetic regions on the surface, and larger magnetic regions at activity 
maximum would produce a stronger rotational modulation of the $S$-index. The significant 
rotation signals ranged from 34.9 to 44.1 days, with one outlier (FAP = 3.3\%) at 15.1 
days in the 1985 season. We reject this period based on the expectation of less extreme 
DR from observations of the Sun and other stellar analyses \citep{Donahue1996}. Indeed, 
with our threshold of FAP $< 5\%$ we expect a spurious detection in 1 of 20 seasons, on 
average, with the FAP defined as the probability of pure noise producing a peak above the 
selected power threshold. Excluding the 1985 season and taking the error-weighted mean of 
the remaining seasonal periods, we find a mean rotation period of $P_{\rm rot} = 40.3 \pm 
1.5$ days, where the uncertainty is the error in the mean, $\sigma/\sqrt{N}$.

The range of the seasonal rotation periods $\Delta P = P_{\rm max} - P_{\rm min}$ is 9.2 
days. This compares well to the work of \cite{Donahue1996} who found a power law 
relationship $\Delta P \propto P^{1.3 \pm 0.1}$ for an ensemble of stars in the MWO 
survey.  Donahue did not publish the proportionality constant, but using their data for 
the Sun ($P_{\rm rot} = 26.09$ days, $\Delta P = 4.0$ days), we find it to be $C = 
0.0576$.  Finally, using $\Delta P_{\rm D96} = C P^{1.3}$ with our mean rotation period 
we obtain the empirical range of 7.0 days, comparable to our measured value of 9.2 days 
(5.0 days if we exclude $P_{\rm rot}=34.9$~days, which has the largest FAP). This 
provides additional confidence that we have found a reasonable spread in rotation periods 
that resembles DR. It is likely a lower limit on the magnitude of DR, as magnetic 
features are not expected to appear at all latitudes.

\cite{Donahue1993} further examined stellar DR by constructing an ``active region 
migration diagram'' consisting of seasonal $P_{\rm rot}$ versus mean seasonal activity 
$\langle S \rangle$.  For the Sun, this produces a wedge-like diagram as the solar cycle 
begins with low activity and long rotation period with spots emerging at the highest 
latitudes, proceeds to high-activity solar maximum with mid-latitude spots and medium 
rotation period, and finishes with low activity again and fast rotation as spots appear 
near the equator.  Donahue proposed that if we assume solar-like active region migration 
from high to low latitudes (an unsupported assumption) then anti-solar DR would manifest 
as a similar wedge-like pattern in opposite time order. Donahue identified several 
candidates for anti-solar DR using this technique. We attempted to apply this technique 
to our data for \aql, but there are not enough sequential seasonal rotation measurements 
and the activity does not have sufficiently long rise/decline phases to apply the method. 
Therefore, while we find that \aql has a range of seasonal rotation periods consistent 
with DR, we cannot determine whether it follows a solar-like or anti-solar pattern from 
the rotation and activity data alone.

\subsection{Spectral Energy Distribution}\label{sec2.6}

To obtain a luminosity constraint for asteroseismic modeling, we performed an analysis of 
the broadband SED of \aql together with the Gaia DR3 parallax \citep[with no systematic 
offset applied; see, e.g.,][]{StassunTorres2021}, following the procedures described in 
\citet{Stassun2016} and \cite{Stassun2017, Stassun2018}. We obtained the $JHK_S$ 
magnitudes from {\it 2MASS}, the $UBV$ magnitudes from the homogeneous photometric 
catalog of \citet{Mermilliod2006}, and the Str\"omgren $uvby$ magnitudes from 
\citet{Paunzen2015}. Together, the available photometry spans the full optical-IR stellar 
SED over the wavelength range 0.3--2~$\mu$m.

We performed a fit using Kurucz stellar atmosphere models, with the effective temperature 
$T_{\rm eff}$, surface gravity $\log g$, and the metallicity [M/H] adopted from 
\cite{Brewer2016}. The extinction $A_V$ was set to zero based on the proximity of the 
system. The resulting fit has a reduced $\chi^2=1.2$. Integrating the model SED gives the 
bolometric flux at Earth, $f_{\rm bol} = 2.243 \pm 0.079 \times 10^{-7}$ 
erg~s$^{-1}$~cm$^{-2}$. Taking the $f_{\rm bol}$ together with the Gaia parallax directly 
yields the bolometric luminosity, $L_{\rm bol} = 1.557 \pm 0.055$~L$_\odot$. The stellar 
radius follows from the Stefan-Boltzmann relation, giving $R = 1.333 \pm 
0.024$~R$_\odot$. In addition, we can estimate the stellar mass from the empirical 
relations of \citet{Torres2010}, giving $M = 1.14 \pm 0.07$~M$_\odot$.

\section{Modeling}\label{sec3}

Below we use the observational constraints from Section~\ref{sec2} to infer additional 
characteristics of \aql from stellar modeling. In Section~\ref{sec3.1} we determine 
precise stellar properties from asteroseismic modeling. In Section~\ref{sec3.2} we 
estimate the wind braking torque from the prescription of \cite{FinleyMatt2018}. Finally, 
in Section~\ref{sec3.3} we constrain the angular momentum history from rotational 
evolution modeling.

\subsection{Asteroseismology}\label{sec3.1}

Using the oscillation frequencies identified in Table~\ref{tab1}, spectroscopic 
parameters from \cite{Brewer2016}, and the luminosity constraint from 
Section~\ref{sec2.6}, five teams inferred the properties of \aql from asteroseismic 
modeling. A variety of stellar evolution codes and fitting methods were used, including 
ASTEC / AMP~1.3 \citep{ChristensenDalsgaard2008a, Metcalfe2009, Creevey2017}, GARSTEC / 
BASTA \citep{Aguirre2022}, MESA / AMP~2.0 \citep{Paxton2015, Metcalfe2023c}, and YREC 
\citep{Demarque2008}. We found reasonable agreement between the results for the 
asteroseismic properties, with individual estimates ranging from 
$R=1.341$--1.397~$R_\odot$, $M=1.02$--1.15~$M_\odot$, and stellar ages between 
8.3--10.1~Gyr. We adopt the values and uncertainties from the AMP~2.0 pipeline, which 
yielded stellar properties that are most representative of the ensemble of results and 
produced an optimal reference model with a reduced $\chi^2 \sim 1$. The adopted 
properties for \aql are summarized in Table~\ref{tab5}.

 \begin{deluxetable}{@{}lcc}
 \setlength{\tabcolsep}{16pt}
 \tablecaption{Adopted Properties of the Subgiant \aql\label{tab5}}
 \tablehead{\colhead{}            & \colhead{\aql}& \colhead{Source}}
 \startdata
 $T_{\rm eff}$ (K)                & $5587 \pm 78$          & 1 \\
 $[$M/H$]$ (dex)                  & $+0.32 \pm 0.07$       & 1 \\
 $\log g$ (dex)                   & $4.15 \pm 0.08$        & 1 \\
 $v \sin i$ (km s$^{-1}$)         & $1.3 \pm 0.5$          & 1 \\
 $B-V$ (mag)                      & $0.77$                 & 2 \\
 $\log R'_{\rm HK}$ (dex)         & $-5.099$               & 2 \\
 Inclination (deg)                & $61$                   & 3 \\
 $|B_{\rm d}|$ (G)                & $8.5 \pm 1.1$          & 3 \\
 $L_X$ ($10^{27}$~erg~s$^{-1}$)   & $2.4 \pm 0.3$          & 4 \\
 Mass-loss rate ($\dot{M}_\odot$) & $1.05^{+1.08}_{-0.58}$ & 4 \\
 $P_{\rm rot}$ (days)             & $40.3 \pm 1.5$         & 5 \\  
 Luminosity ($L_\odot$)           & $1.557 \pm 0.055$      & 6 \\
 Radius ($R_\odot$)               & $1.358 \pm 0.016$      & 7 \\
 Mass ($M_\odot$)                 & $1.07 \pm 0.04$        & 7 \\
 Age (Gyr)                        & $8.9 \pm 1.3$          & 7 \\
 \hline
 Torque ($10^{30}$~erg)           & $3.22^{+2.52}_{-1.53}$ & 8 \\
 Rossby number ($\Rosun$)         & $1.29 \pm 0.06$        & 9 \\
 \enddata
 \tablerefs{(1)~\cite{Brewer2016}; (2)~\cite{Baliunas1996}; 
  (3)~Section\,\ref{sec2.2}; (4)~Section\,\ref{sec2.3}; (5)~Section\,\ref{sec2.5}; 
  (6)~Section\,\ref{sec2.6}; (7)~Section\,\ref{sec3.1}; (8)~Section\,\ref{sec3.2}; 
  (9)~Section\,\ref{sec3.3}}
 \vspace*{-24pt}
 \end{deluxetable}
 \vspace*{-12pt}

The AMP~2.0 pipeline \citep{Metcalfe2023c} uses a parallel genetic algorithm 
\citep{Metcalfe2003} to optimize the match between the observational constraints and 
stellar models produced by version r12778 of the MESA stellar evolution code 
\citep{Paxton2015} and version 6 of the GYRE pulsation code \citep{Townsend2013}.  The 
stellar models use the default MESA equation of state, which is primarily from OPAL 
\citep{Rogers2002} and SCVH \citep{Saumon1995}, OPAL opacities supplemented with 
low-temperature values from \cite{Ferguson2005}, the \cite{Grevesse1998} solar mixture, 
NACRE nuclear reaction rates \citep{Angulo1999}, a gray atmosphere with 
\cite{Eddington1926} $T$-$\tau$ integration, the mixing-length formalism of 
\cite{Cox1968}, diffusion and settling of helium and heavy elements from 
\cite{Thoul1994}, and the two term \cite{Ball2014} prescription for the asteroseismic 
surface correction. The values and uncertainties of the asteroseismic properties were 
determined from the likelihood-weighted mean and standard deviation of all models sampled 
by the genetic algorithm during the optimization.

\subsection{Magnetic Evolution}\label{sec3.2}

We can now estimate the current wind braking torque for \aql using the prescription of 
\cite{FinleyMatt2018}. Combining the large-scale magnetic field strength derived from 
spectropolarimetry in Section~\ref{sec2.2}, the mass-loss rate inferred from the \xray 
surface flux in Section~\ref{sec2.3}, the mean rotation period from the chromospheric 
activity data in Section~\ref{sec2.5}, and the precise stellar mass and radius from 
asteroseismology in Section~\ref{sec3.1}, our wind braking torque estimate is 
$3.22^{+2.52}_{-1.53} \times 10^{30}$~erg. The dominant contributions to the total torque 
uncertainty ($+78\%, -48\%$) come from the errors on the estimated mass-loss rate 
($+47\%, -35\%$) and the dipole field strength ($\pm12\%$), followed by smaller 
contributions from the rotation period ($\pm4\%$), the stellar radius ($\pm4\%$), and the 
stellar mass ($\pm1\%$). Similar estimates of the wind braking torque using the values of 
$B_{\rm q}$ or $B_{\rm o}$ derived in Section~\ref{sec2.2} yield $3.08^{+2.77}_{-1.58} 
\times 10^{30}$~erg (quadrupole) or $1.10^{+1.21}_{-0.63} \times 10^{30}$~erg (octupole), 
consistent with the range of values found from $B_{\rm d}$.

The wind braking torque for \aql is nearly an order of magnitude stronger than that of 
the younger solar analog 16\,Cyg\,A \citep[cf.][]{Metcalfe2022}. The two stars have 
nearly the same mass and chromospheric activity, but \aql has a lower surface gravity, 
larger radius, and slightly higher luminosity than 16\,Cyg\,A, as well as an 
asteroseismic age that is about 1.5~Gyr older. Despite the difference in metallicity, if 
we treat these two stars as an evolutionary sequence we can investigate the sources of 
the large difference in their wind braking torques by changing one parameter at a time 
between the fiducial models for each star. We find that the large increase in the wind 
braking torque in this age range (7.4--8.9 Gyr) is driven primarily by the much stronger 
dipole field ($+1240\%$), along with the slightly larger radius ($+39\%$) and higher 
mass-loss rate ($+7\%$) of \aql, offset by the substantially longer rotation period 
($-49\%$).

Considering the simulation results of \cite{Noraz2024}, it's possible that \aql may have 
reached the higher Ro regime where DR is predicted to shift from solar-like (fast 
equator, slow poles) to anti-solar (slow equator, fast poles). This transition apparently 
leads to a restoration of the large-scale magnetic field on the subgiant branch, after a 
prolonged phase of WMB during the second half of the main-sequence lifetime. Unlike the 
large-scale fields of younger main-sequence stars, the field produced by anti-solar DR in 
the simulations is stationary rather than cycling---in agreement with the flat activity 
record shown in Figure~\ref{fig4}. We hypothesize that the restoration of the large-scale 
magnetic field with the transition to anti-solar DR produces a phase of revived magnetic 
braking on the subgiant branch. We explore this scenario using a toy model in the 
following section.

\subsection{Rotational Evolution}\label{sec3.3}

If we model \aql with assumptions identical to those in \citet{Metcalfe2025b} but 
including an asteroseismic prior on the stellar age (a Gaussian centered at $8.9$~Gyr 
with $\sigma=1.3$~Gyr), we predict a rotation period $P_{\rm rot}=33\pm4$~days under WMB, 
and $P_{\rm rot}=58\pm9$~days under a standard spin-down scenario---both in mild tension 
with the observed rotation period. \aql appears to be a contradiction: it has an observed 
instantaneous torque consistent with standard braking, but it cannot have undergone that 
level of spin-down for its entire life and still retain its relatively rapid rotation. 
The simulations of \citet{Noraz2024} suggest a possible means to reconcile these 
observations: the establishment of anti-solar DR may eventually restore the torque.

 \begin{figure}[!t]
 \centering\includegraphics[width=\columnwidth, trim=7 7 5 5,clip]{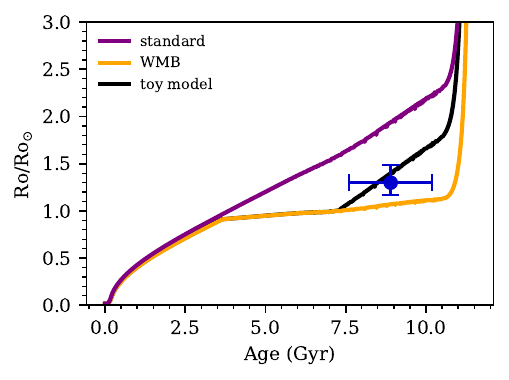} 
 \caption{Rossby number as a function of time for a standard spin-down model (purple), 
WMB model (orange), and a toy model in which braking ceases at $\Rocrit$ and resumes at 
$\Roflip$ (black). The standard and WMB cases are the median posterior tracks for models 
constrained to match the surface temperature, radius, and metallicity with an informative 
asteroseismic age prior. The black curve adds the parameter $\Roflip$ by including the 
observed rotation period as a constraint on the model.\label{fig5}}
 \end{figure}

To examine this scenario, we construct a toy model that incorporates a phase of WMB that 
begins at $\Rocrit$, with revived magnetic braking at a higher Ro where DR is thought to 
flip from solar-like to anti-solar ($\Roflip$). The model adopts solid body rotation, 
although the DR profile may influence the torque \citep[e.g.,][]{Tokuno2023, Finley2023}. 
In this toy model, both transitions are abrupt: angular momentum loss is set to zero at 
$\Rocrit$, and set to the standard braking after $\Roflip$, given the instantaneous 
rotation period and stellar structure. In between these two transitions, the rotational 
evolution is dictated solely by the evolving moment of inertia as the star gradually 
expands on the main-sequence. Despite the lack of a wind braking torque, Ro slowly 
increases with time during this phase (see the orange line in Figure~\ref{fig5}).

We fit for a value of $\Roflip$ using the following procedure. We fix the values of 
$\Rocrit$ and the braking normalization ($f_k$) to those determined in 
\citet{Saunders2024} for WMB, but we leave $\Roflip$ as a free parameter (along with the 
model mass, metallicity, age, and mixing length). The prior on $\Roflip$ is uninformative 
aside from two limits: we require that $\Roflip > \Rocrit$ and that the stellar age at 
$\Roflip$ is less than the current age, which effectively incorporates a prior from the 
observed wind braking torque for \aql.

We include the rotation period as a constraint on the model, adopting the value from 
Section~\ref{sec2.5} with an additional 10\% error on the period, added in quadrature, to 
account for systematic uncertainty in the observed active latitudes 
\citep[i.e.][]{Epstein2014}. As in \citet{Metcalfe2025b}, the observed $T_{\rm eff}$, 
[M/H], and radius are used as constraints on the model, with values and errors adopted 
from Table \ref{tab5}, along with additional systematic uncertainties added in 
quadrature.

We find $\Roflip = 1.1\pm0.05\ \Rocrit$. Because the evolution between $\Rocrit$ and 
$\Roflip$ is driven by slow structural changes to the star over a nuclear timescale, 
there is a sizable delay before the star reaches $\Roflip$---in our model for \aql, 
braking resumes $3.5^{+1.3}_{-2.2}$~Gyr after the onset of WMB, at an age of 
$7.1^{+1.6}_{-2.5}$~Gyr. This represents a substantial fraction of the main-sequence 
lifetime. The long pause in spin-down means that our toy model preserves many of the 
broad features of the original WMB model: stars undergo a prolonged phase of subdued 
spin-down during the latter half of the main-sequence, and thus a sizable population of 
objects should be observed in the midst of WMB.

\section{Discussion}\label{sec4}

\begin{figure}[!t]
 \centering\includegraphics[width=\columnwidth]{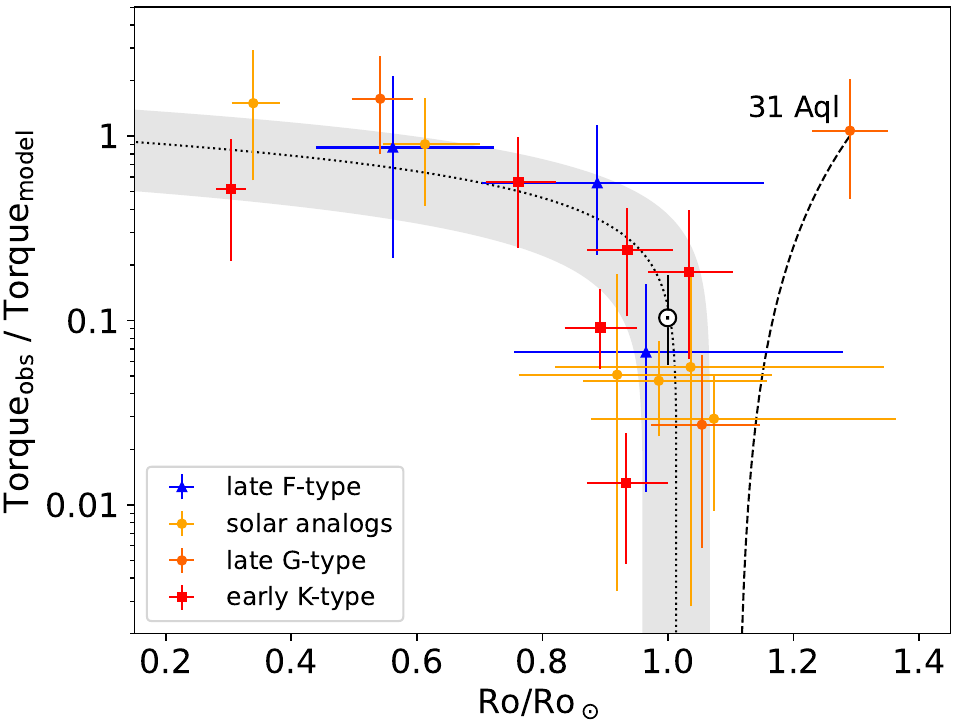} 
 \caption{Ratio of the observationally estimated and model wind braking torque as a 
function of Rossby number normalized to the solar value. The dotted line and gray shaded 
region illustrates the fit and 95\% confidence interval described by 
\cite{Metcalfe2025b}. The revived magnetic braking in \aql at higher Ro is evident, and 
quadratic growth is illustrated schematically with a dashed line.\label{fig6}}
 \end{figure}

The estimated wind braking torque and current Ro of \aql appear to constrain the 
transition from solar-like to anti-solar DR in subgiant stars. In Figure~\ref{fig6} we 
reproduce the evidence from \cite{Metcalfe2025b} for the gradual onset of WMB in older 
main-sequence stars, to which we add the new constraint from the G7 subgiant \aql 
(labeled orange point). The efficiency of magnetic braking decreases by up to two orders 
of magnitude as stars approach $\Rocrit$ slightly above the solar value. To explain the 
current wind braking torque and rotation period of \aql, we find that standard magnetic 
braking may resume near 1.1~$\Rocrit$---apparently from a restoration of the large-scale 
magnetic field due to a shift of the DR pattern from solar-like to anti-solar, as 
suggested by recent simulations \citep{Noraz2024}.

Although we cannot directly constrain the sense of the DR from the chromospheric activity 
data, the observations reveal a variety of rotation periods in different seasons and 
minimal long-term variability on cycle timescales, consistent with the simulations. 
Previous support for the simulation results came from observations of enhanced 
photometric variability in Kepler targets at Ro above the solar value \citep{Mathur2025}, 
but \aql suggests that the transition to anti-solar DR might be accompanied by a 
restoration of the large-scale magnetic field and revived magnetic braking.

The revived magnetic braking due to anti-solar DR is distinct from the born-again dynamo 
phenomenon discovered in 94\,Aqr\,Aa \citep{Metcalfe2020} and confirmed in $\beta$\,Hyi 
\citep{Metcalfe2024b, Santos2025}. The latter occurs in stars with shallower convective 
envelopes, when physical expansion on the subgiant branch initially pushes Ro above 
$\Rocrit$, which is thought to lead to a shutdown of the cycling global dynamo. However, 
subsequent structural evolution pushes Ro back below $\Rocrit$, apparently reinvigorating 
the dynamo near the base of the red giant branch. Stars with deeper convection zones like 
\aql do not experience a ``blue hook'' in the H-R diagram, but instead stellar evolution 
models show a steadily increasing Ro along the subgiant branch. This steady evolution 
apparently leads to a shutdown of the cycling global dynamo when Ro exceeds $\Rocrit$, 
followed by revived magnetic braking when Ro exceeds $\Roflip$ and a non-cycling 
large-scale magnetic field is produced. Despite these distinct trajectories, it is 
possible that stars with born-again dynamos might have exceeded $\Roflip$ somewhere along 
their evolutionary paths, which may provide complementary constraints on $\Roflip$.

A more sophisticated treatment of the rotational evolution would incorporate the gradual 
onset of WMB as Ro approaches $\Rocrit$ and the gradual resumption of magnetic braking 
above $\Roflip$. Our toy model is a step function that eliminates magnetic braking at 
$\Rocrit$ and another step function that resumes magnetic braking at $\Roflip$. The 
functional form of the gradual onset shown in Figure~\ref{fig6} is a generic prediction 
of a Hopf bifurcation, with the dynamo efficiency decreasing as $\sqrt{\Rocrit-{\rm Ro}}$ 
\citep{Cameron2017}. This treatment is appropriate for the transition from an oscillatory 
to a stationary solution, while the shift from solar-like to anti-solar DR in a 
non-oscillatory dynamo can be described by a transcritical bifurcation with a quadratic 
growth in dynamo efficiency at ${\rm Ro} > \Roflip$ (dashed line in Figure~\ref{fig6}). A 
refined analysis that replaces the step functions in our toy model with the expected 
functional forms for these two transitions could improve our estimate of the absolute 
value of Ro (not just relative to $\Rocrit$) where DR shifts from solar-like to 
anti-solar.

It is important to note that there are still significant uncertainties, in both the 
observations and the simulations, about the exact timing of the transition from 
solar-like to anti-solar DR. The value of $\Rocrit \sim 1.01\ \Rosun$ found by 
\cite{Metcalfe2025b} approximately separates stars with cycles at lower Ro and stars with 
flat activity at higher Ro. This suggests that Ro can exceed $\Rocrit$ slightly during 
the latter half of the main-sequence lifetime, due to mild physical expansion even when 
magnetic braking is substantially weakened. This expansion accelerates during the 
subgiant phase, and our fit to \aql suggests a value of $\Roflip \sim 1.1\ \Rocrit$ that 
is significantly higher than stars typically reach on the main-sequence. However, with 
only one star currently available to establish this constraint, the transition to 
anti-solar DR may be possible at lower values of Ro.

The simulations described by \cite{Brun2022} and analyzed by \cite{Noraz2024} were not 
designed to probe evolution beyond the main-sequence, so their constraints on $\Roflip$ 
may not be entirely correct for a subgiant. Although the authors noted the expected 
influence of WMB, the simulations attempted to follow the evolution of main-sequence 
stars under a scenario of standard magnetic braking. As such, the background stellar 
structure in the simulations at Ro above the solar value was appropriate for evolved 
main-sequence stars rather than subgiants \citep[see also][]{Hotta2026}. The limited 
sampling of Ro in these computationally expensive simulations currently suggests the 
existence of $\Roflip$ somewhere between \mbox{0.76--1.38~$\Rosun$} \citep[see Fig.1 
in][]{Noraz2024}. Future simulations in this range of Ro with a background stellar 
structure appropriate for subgiant stars may provide stronger constraints on the value of 
$\Roflip$.

The inclination of \aql ($i\sim61^\circ$) is favorable for the asteroseismic detection of 
rotational splittings that could distinguish between solar-like and anti-solar DR. In 
particular, the visibility of both the inner ($m=\pm1$) and outer ($m=\pm2$) splittings 
for an $\ell=2$ quintuplet is optimal at $i=60^\circ$ \citep{Gizon2003}. The latter 
($\ell=m$) probe the equatorial rotation rate while the former ($\ell \ne m$) probe the 
rotation at higher latitudes \citep{Lund2014}, so the relative splittings can reveal the 
sense of the DR. This technique has already been demonstrated for several dozen Kepler 
targets \citep{Benomar2018, Bazot2019}, requiring long-term high S/N observations. Such 
measurements may be possible from a careful analysis of KIC\,8349582 (Kepler-95), a close 
analog of \aql in the Kepler field of view. The duration of TESS observations is 
currently too limited to resolve rotational splittings, and the proposed PLATO fields do 
not include \aql \citep{Nascimbeni2022}. However, the higher S/N of ground-based radial 
velocity observations for asteroseismology should be able to resolve the rotational 
splittings and decipher between solar-like and anti-solar DR in \aql. Future multi-month 
monitoring by the Stellar Observations Network Group \citep[SONG;][]{Grundahl2008} might 
provide the necessary S/N for such a measurement.

\vspace*{12pt}
T.S.M.\ acknowledges support from Chandra award GO0-21005X, NSF grant AST-2205919, TESS General Investigator grant 80NSSC25K7898, and NASA grant 80NSSC25K7563. Computational time at the Texas Advanced Computing Center was provided through ACCESS allocations TG-AST090107 and TG-PHY250037. 
J.v.S.\ acknowledges support from NSF grant AST-2205888.
D.B.\ acknowledges support from TESS General Investigator grant 80NSSC19K0385.
R.A.G.\ acknowledges support from the GOLF and PLATO Centre National D'{\'{E}}tudes Spatiales (CNES) grant.
O.K.\ acknowledges support from the Swedish Research Council (grant agreement no. 2023-03667) and the Swedish National Space Agency.
S.H.S.\ is grateful for support from award HST-GO-15991.002-A.
A.S.\ acknowledges support from the European Research Council (ERC) under the European Union's Horizon 2020 research and innovation program (CartographY; grant agreement 804752). 
T.R.B.\ is supported by the Australian Research Council (FL220100117).
S.N.B.\ acknowledges support from PLATO ASI-INAF agreement no.\ 2022-28-HH.0 ``PLATO Fase D''.
The TESS and HST data used in this paper can be found on MAST: \dataset[https://doi.org/10.17909/0vak-d288]{https://doi.org/10.17909/0vak-d288}.


\end{document}